\documentclass[mathleft
]{an}
\usepackage{graphicx}
\usepackage{times}
\begin{document}

\Pagespan{1}{3}
\Yearpublication{2012}%
\Yearsubmission{2012}%
\Month{11}%
\Volume{999}%
\Issue{88}%

\title{New pulsating variable stars in the Stock~14 open cluster and surrounding fields}

\author{D. Drobek\thanks{\email{drobek@astro.uni.wroc.pl}\newline} }

\titlerunning{New pulsating variable stars in the Stock~14 open cluster and surrounding fields}
\authorrunning{D. Drobek}
\institute{
Instytut Astronomiczny Uniwersytetu Wroc{\l}awskiego, Kopernika 11, 51-622 Wroc{\l}aw, Poland
}

\received{October 2012}
\accepted{October 2012}
\publonline{November 2012}


\abstract{This paper presents the preliminary results of a~multicolour photometric study of the young open cluster Stock~14 and
adjacent fields. The reddening, distance and age of the cluster were determined from colour-colour and colour-magnitude
diagrams by means of isochrone fitting. Fourier analysis of the acquired time-series data was performed, which resulted in the
discovery of new pulsating variable stars and candidates for such objects.}

\maketitle

\section{Introduction}

The young galactic open cluster Stock~14 is known to host two eclipsing binary systems with $\beta$~Cephei-type pulsating
components, HD~101497 and HD~101838 (Pigulski \&~Poj\-ma\'{n}\-ski 2008). After the discovery of those systems the cluster was
subject to a~photometric follow-up study in order to confirm the finding (Drobek et al. 2010). The field of view of the telescope
used for this follow-up encompassed the entire cluster and some adjacent regions of the sky. As a~consequence, $UBV$ time-series
data of many objects in vicinity of the two program stars were acquired. This data set was well suited for a~comprehensive
variability survey in the observed field. No such study of the Stock~14 cluster has been conducted before. Because the observed
field is located in the vicinity of the Galactic plane ($b \approx -0.7^{\circ}$) and covers a~part of the Cru~OB1 association,
one can expect to discover pulsating stars of early spectral types.

\section{Observing run and data reduction}

The photometric data used in this study were acquired by R.\,Shobbrook and A.\,Narwid during 19 nights between March and May 2007
with the 1.0-m telescope in Siding Spring Observatory, Australia. The detector being used was the Wide Field Imager, with a~field
of view of about 0.5 square degrees. The frames were calibrated using a~standard procedure which consisted of overscan subtraction
and trimming, bias subtraction, linearity correction and flat-fielding. Almost 25\,000 stars were identified in the $V$-filter
reference frame. The stellar magnitudes were calculated with the Daophot~II software package (Stetson 1987). In order to remove
the atmospheric effects from the resulting time series, differential photometry was used comparing to a~set of stars which
were evenly distributed in the observed field. The instrumental $UBV$ magnitudes of each star were calculated as averages of its
$\sigma$-clipped time series. Because no photometric standards were observed during the run, the transformation of
instrumental magnitudes to the standard system was based on the photoelectric data of Peterson \&~FitzGerald (1988).

\section{Cluster parameters}

In this study, the number of stars with measured $UBV$ co\-lour indices is greater than in the works of Moffat \&~Vogt (1975),
Turner (1985), Peterson \&~FitzGerald (1998) and Kharchenko et al. (2005) who previously determined the parameters of Stock~14.
For this reason, an attempt to derive better values of cluster parameters was made. In order to reduce the number of field
interlopers, a~subset of stars located within 10 arc minutes from the centre of the cluster ($\alpha_{\rm 2000} =$
11$^{\rm{h}}$43.6$^{\rm{m}}$, $\delta_{\rm 2000} = -$62$^{\circ}$31$^\prime$) was selected.

\begin{figure}
    \includegraphics[width=83mm]{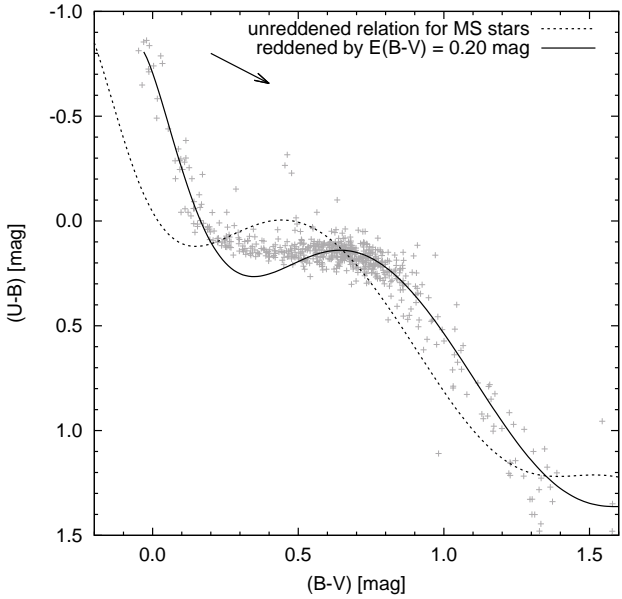}
    \caption{Determination of the reddening of the Stock~14 cluster. Dashed line represents the unreddened relation, solid
line represents the same relation reddened by $E(B-V)=$ 0.20 mag. The arrow is a~reddening vector corresponding to the derived
colour excess.}
\label{caldwell}
\end{figure}

The reddening of Stock~14 was estimated by fitting the relation between the intrinsic colours of the main sequence stars
(Caldwell et al. 1993) to the colour-colour dia\-gram of the cluster (Fig.~\ref{caldwell}). A~standard relation between colour
excesses, $E(U-B)/E(B-V)=$ 0.72 was assumed. The reddening $E(B-V)=$ 0.20 $\pm$ 0.02 mag was derived. It should be noted that the
reddened intrinsic colour relation does not fit equally well to all the stars in the plot. There is a~discrepancy for stars with
$(B-V)$ in the range from 0.2 to 0.5~mag. It seems to be caused by the difference between the standard and the instrumental $U$
filters. This causes the $(U-B)$ colours to be unreliable for some stars. However, this does not affect the determination of
reddening, as the intrinsic relation still fits the data very well in the regions occupied by stars of early and late spectral
types.

\begin{figure}
    \includegraphics[width=83mm]{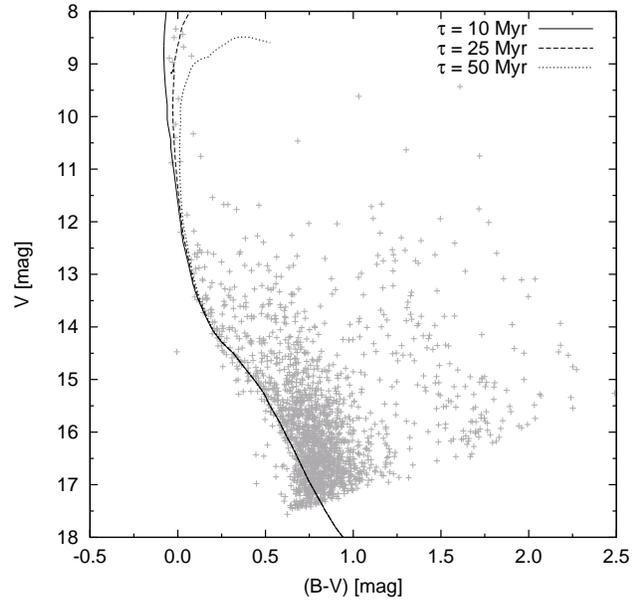}
    \caption{Determination of the age and distance to the Stock~14 cluster. The best-fitting isochrone corresponds to the true
distance modulus $(m-M)_0=$~11.63~mag and the age $\tau=$~25~Myr. Solar metallicity was assumed, and the effects of rotation were
not taken into account.}
\label{isofit}
\end{figure}

The colour-magnitude diagram of the stars in the vici\-ni\-ty of the cluster (Figs.~\ref{isofit} and \ref{cmd_inside}) shows
a~well-defined main sequence down to $V$ magnitude of about 15.5~mag. Once the reddening of Stock~14 was determined, the
isochrones of Ekstr\"{o}m et al. (2012) were fitted to the main sequence of the cluster in the colour-magnitude plane. Solar
metalli\-city ($Z=$ 0.014) was assumed. The results of fitting were si\-mi\-lar regardless of including or neglecting the effects
of stellar rotation on the isochrones. The true distance modulus was estimated at $(m-M)_0=$~11.63 $\pm$ 0.05~mag. The cluster
seems to be between 10 and 50~Myr of age, with the best-fitting isochrone corresponding to $\tau=$~25~Myr. Since the pre\-vious
estimates of the parameters of Stock~14 were based on smaller samples of stars, the results presented here should be regarded as
more reliable.

\section{Variable stars}

With up to 1200 data points per star, the $V$-filter time series data were best suited for the variability survey. A~discrete
Fourier transform of every time series was calculated in the frequency range from 0 to 40 d$^{-1}$. In the re\-sul\-ting Fourier
spectra the maxima with the signal-to-noise ratio above 4.0 were assumed to be significant. All of the re\-sul\-ting frequency
spectra and light curves were examined by eye, and classified based on the available information, such as the shape of the light
curve, frequencies, their corres\-ponding light amplitudes, $UBV$ colours, or spectral types found in the literature.

\begin{table}
\centering
\caption{Summary of the discovered multiperiodic and monoperiodic pulsating variable stars and candidates.}
\label{tabvar}
\begin{tabular}{ccc}
    \hline
    Variability & Multiperiodic & Monoperiodic\\
    type & (pulsating) & (candidates) \\
    \hline
    $\beta$~Cep & 2 & $-$ \\
    SPB & 2 & 5\\
    $\delta$~Sct & 8 & 9 \\
    $\gamma$~Dor / $\delta$~Sct & 2 & $-$ \\
    $\gamma$~Dor & 5 & $-$ \\
    unknown & 6 & 5 \\
    \hline
\end{tabular}
\end{table}

Out of the 92 new variable objects, 25 exhibit multi\-pe\-riodic sinusoidal light variations. Detection of multiple independent
frequencies usually indicates that light variations are caused by stellar pulsation. For this reason, all the discovered
mul\-ti\-pe\-rio\-dic stars are assumed to be pulsating. Such assumption is not valid for the 19 mono\-perio\-dic va\-riab\-les,
which are con\-si\-dered candidates for pulsating stars. Alternative explanations such as stellar rotation or ellip\-soidal
variability have to be taken into account. A~summary of the numbers of discovered (candidate) pulsating stars and their
corresponding varia\-bility classes is presented in Table~\ref{tabvar}.

\begin{figure}
    \includegraphics[width=83mm]{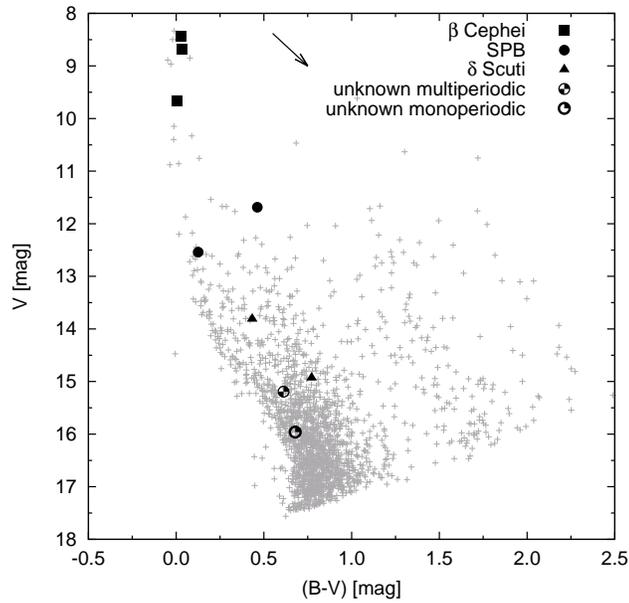}
    \caption{Colour-magnitude diagram of the stars located up to 10 arc minutes from the centre of Stock~14. The
symbols denote the discovered variable stars. The arrow is a~reddening vector corresponding to $E(B-V)=$~0.20~mag.}
\label{cmd_inside}
\end{figure}

\begin{figure}
    \includegraphics[width=83mm]{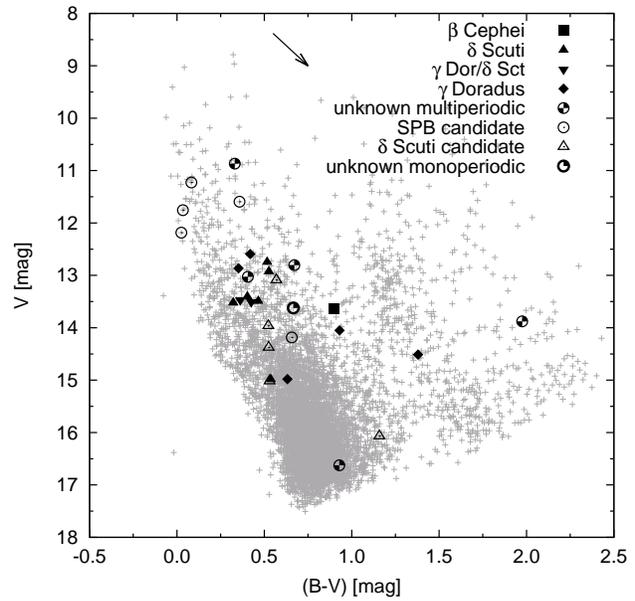}
    \caption{Colour-magnitude diagram of the stars located further than 10 arc minutes from the centre of Stock~14. The meaning of
symbols is similar to Fig.~\ref{cmd_inside}. The vast majority of the stars in this diagram are field objects.}
\label{cmd_outside}
\end{figure}

The positions of the discovered variable stars in the colour-magnitude diagrams are presented in two figures. In
Fig.~\ref{cmd_inside}, only the stars located within 10 arc minutes from the centre of Stock~14 have been plotted. Because
of the resultant reduction of the number of field stars, it is possible to distinguish a~well-defined main sequence of the
cluster. The two brightest $\beta$~Cephei-type variables are the previously known pulsating stars in eclipsing binary systems,
HD~101794 and HD~101838. They are shown here for the sake of completeness. In Fig.~\ref{cmd_outside}, the remaining stars from the
observed field are shown. This plotted subset of stars is a~mixture of objets located at various distances from the Sun and
reddened in a~non-uniform manner. This can explain somewhat surprising positions of several variable stars in the colour-magnitude
plane such as the $\beta$~Cephei-type variable and one of the SPB candidates. Several of the 44 stars listed in Table~\ref{tabvar}
were very faint in $B$ filter, which made their $(B-V)$ colour indices unreliable. Those stars are not shown in the
colour-magnitude diagrams.

A~detailed analysis of the properties of the discovered variable stars and their membership to Stock~14 is ongoing and will be
published elsewhere. However, even after a~cursory glance at the data it is safe to assume that the majority of the found variable
objects are field stars not physically associated with the cluster.

\section{Conclusions}

The reddening and distance estimates show smaller reddening and smaller distance to Stock~14 in comparison to the results of
Moffat \&~Vogt (1975), Turner (1985) and Peterson \&~FitzGerald (1998). On the other hand, they are in good agreement with the
results of Kharchenko et al. (2005). The difference is likely caused by using a~different intrinsic colour relation for main
sequence stars as well as better quality of the photometric data and of the model isochrones.

Discovery of many pulsating stars and candidates for such objects is not surprising, taking into account that a~dense field near
the Galactic plane was observed. The majority of the new variable objects seem to be field stars not physically associated with
Stock~14.

\acknowledgements
The author wishes to thank the Director of the Research School of Astronomy and Astrophysics of the Australian National
University for time on the 40-inch telescope at Siding Spring Observatory. This research has made use of the WEBDA database,
operated at the Institute for Astronomy of the University of Vienna, the Vizier catalogue access tool, CDS, Strasbourg, France and
the SAOImage DS9, developed by Smithsonian Astrophysical Observatory. This work was supported by the National Science Centre
grant No. 2011/01/N/ST9/00400.

\end{document}